\documentclass[twocolumn]{aastex63}
\usepackage{multirow}

\received{April 18, 2022}
\revised{April 18, 2022}
\accepted{\today}
\usepackage{placeins}
\FloatBarrier
\submitjournal{AJ}
\usepackage{float}
\usepackage[ansinew]{inputenc}
\usepackage[american]{babel}
\usepackage{etex}
\usepackage{stackengine}
\usepackage{csquotes}
\usepackage{ifthen}
\usepackage{url}
\usepackage{euscript}
\usepackage{multirow}
\usepackage{nameref}
\usepackage{xcolor}
\usepackage{varioref}
\usepackage{hyperref}
\usepackage{cleveref}

\shortauthors{C. Swastik et al.}

\graphicspath{{./}{figures/}}

\usepackage{float}
\usepackage{natbib}
\begin{document}

\title {Age analysis of extrasolar planets:  Insight from stellar isochrone models }
%\title {Chronology of planet formation: Trends  from stellar isochrone models} 
%\title {Timeline of planet formation: Trends  \& evidence from stellar isochrone models} 
%\title{Formation timeline of exoplanets: Insight from stellar isochrone models}        

\correspondingauthor{C. Swastik}
\email{swastik.chowbay@iiap.res.in}

\author[0000-0003-1371-8890]{C. Swastik}
\affiliation{Indian Institute of Astrophysics, Koramangala 2nd Block, Bangalore 560034, India}
\affiliation{Pondicherry University, R.V. Nagar, Kalapet, 605014, Puducherry, India}

\author[0000-0003-0799-969X]{Ravinder K. Banyal}
\affiliation{Indian Institute of Astrophysics, Koramangala 2nd Block, Bangalore 560034, India}

\author[0000-0002-0554-1151]{Mayank Narang}
\affiliation{Department of Astronomy and Astrophysics, Tata Institute of Fundamental Research
Homi Bhabha Road, Colaba, Mumbai 400005, India}
\affiliation{Academia Sinica Institute of Astronomy $\&$ Astrophysics, 11F of Astro-Math Bldg., No. 1, Sec. 4, Roosevelt Road, Taipei 10617, Taiwan, Republic of China}

\author[0000-0001-6093-5455]{Athira Unni}
\affiliation{Indian Institute of Astrophysics, Koramangala 2nd Block, Bangalore 560034, India}
\affiliation{Department of Physics and Astronomy, University of California, Irvine, 4129 Frederick Reines Hall, Irvine, CA 92697, USA}

\author[0000-0003-0891-8994]{T. Sivarani}
\affiliation{Indian Institute of Astrophysics, Koramangala 2nd Block, Bangalore 560034, India}

% \author[0000-0002-3530-304X]{P. Manoj}
% \affiliation{Department of Astronomy and Astrophysics, Tata Institute of Fundamental Research
% Homi Bhabha Road, Colaba, Mumbai 400005, India}

% \author[0000-0001-8075-3819]{Bihan Banerjee}
% \affiliation{Department of Astronomy and Astrophysics, Tata Institute of Fundamental Research
% Homi Bhabha Road, Colaba, Mumbai 400005, India}

% \author[0000-0003-0003-4561]{S. P. Rajaguru}
% \affiliation{Indian Institute of Astrophysics, Koramangala 2nd Block, Bangalore 560034, India}

\begin{abstract}
There is growing evidence from stellar kinematics and galactic chemical evolution (GCE)  suggesting that giant planets (M$_{P}\geq$0.3$M_{J}$) are relatively young compared to the most commonly occurring population of small planets (M$_{P} <$0.3$M_{J}$). To further test the validity of these results, we analyzed the ages for a large number of 2336 exoplanet hosting stars determined using three different but well-established isochrone fitting models, namely, PARSEC, MIST, and Yonsei Yale (YY). As input parameters, we used Gaia DR3 parallaxes, magnitudes, and photometric temperature, as well as spectroscopically determined more accurate temperatures and metallicities from the Sweet Catalog. Our analysis suggests that  $\sim$~50$\%$ to 70$\%$  of stars with planets are younger than the sun. We also find that, among the confirmed exoplanetary systems, stars hosting giant planets are even younger compared to small planet hosts. The median age of $\sim$~2.61 to 3.48~Gyr estimated for the giant planet-hosting stars (depending on the model input parameters) suggests that the later chemical enrichment of the galaxy by the iron-peak elements, largely produced from Type Ia supernovae, may have paved the way for the formation of gas giants. Furthermore, within the giant planet population itself, stars hosting hot Jupiters (orbital period $\le$10 days) are found to be  younger compared to the stellar hosts of cool and warm Jupiters (orbital period $>$10 days), implying that hot Jupiters could be the youngest systems to emerge in the progression of planet formation. 
\end{abstract}.

\keywords{Stellar ages(1581), Exoplanets(498), Planet hosting stars(1242), Exoplanet formation(492), Extrasolar gaseous giant planets(509), Hot Jupiters(753) }

\section{Introduction}
The field of exoplanets has seen a rapid rise in the past three decades after the discovery of the first planet by \cite{1995Natur.378..355M}. We realized that planets are ubiquitous and their architecture and properties are considerably more diverse and complex  \citep{2012Sci...337..556C,2017Natur.546..514G,2017Natur.542..456G,2021ARA&A..59..291Z}. Apart from the strange planet discovered around the pulsar: PSR1257+12 \citep{1992Natur.355..145W}, the earliest exoplanets discovered were all giants, equivalent to Jupiter in mass and size. Among those initial discoveries were hot Jupiters which orbit their stars with periods of a few days as they were easier to detect compared to smaller planets\citep{1995Natur.378..355M,2000ApJ...529L..45C,2000ApJ...529L..41H}. Thus, the first exoplanet detection, is itself a surprise to the astronomy community since no such planet exists in our solar system. Numerous revelations have emerged over the years since the first discovery. To date, more than 5000 exoplanets have been discovered, which span a wide range of population. While some planets orbit in close proximity to their parent star \citep{2016Natur.533..509M,2016AJ....152..105M,2018MNRAS.474.5523S,2018A&A...612A..95B,2018AJ....155..107M,2021Sci...374.1271L}, others have incredibly extended orbits \citep{2001A&A...375L..27N,2008A&A...480L..33T,2018arXiv181002031C,2021A&A...651A..72V,2021A&A...648A..26W,2022arXiv220505696C,2021ApJ...916L...2B,2023A&A...680A.114R,wahhaj2024pds}. 

Several correlations connecting the stellar and planetary properties have emerged in the past decade \citep{san04,fis05,fis14,nar18,2021A&A...656A..53S,2021AJ....161..114S,2022AJ....164..181U}. One such correlation is the stellar age-planetary mass correlation. The ages of Stars With Planets (SWP) are crucial for investigating numerous aspects of planetary system evolution, such as dynamical interactions among planets \citep{Laughlin_2002} and tidal effects generated by SWP \citep{2004A&A...427.1075P,2009MNRAS.395.2268B}. In fact, understanding the ages of stars holds significant importance in the process of selecting stellar candidates for planet detection \citep{2015A&A...575A..18B} and assessing their potential habitability. The rotation and activity levels of a star which serve as indicators of stellar age, play a crucial role in determining the habitability of planets orbiting around them.

\begin{figure}
\centering
\includegraphics[width=1\columnwidth]{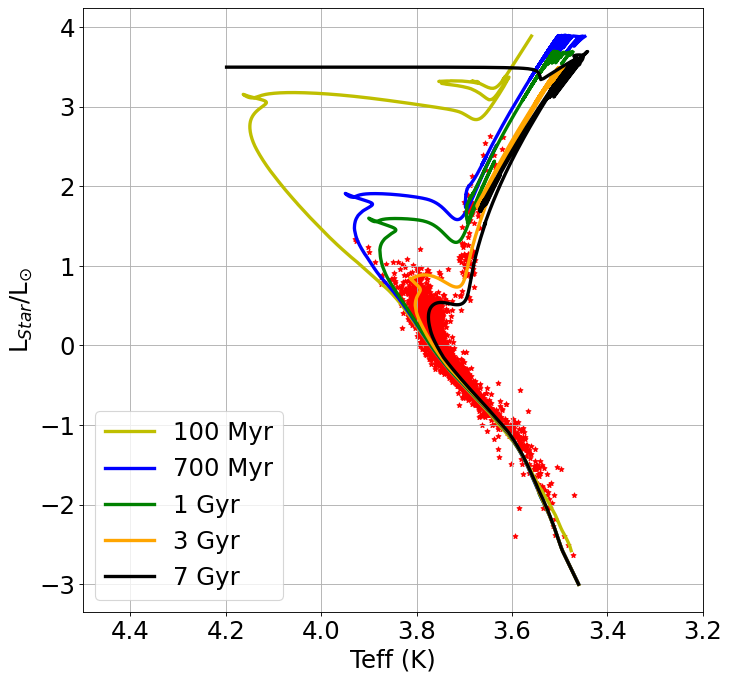}
\caption{Stellar isochrone models generated using MIST. The red symbols represent planet-hosting stars. The isochrones are drawn for the solar-scaled abundances. Both X and Y scales are in logarithmic units.}
\label{f11}
\end{figure}

The majority of planets have been detected around main-sequence FGK stars. Due to the degeneracy of parameters and the slow evolution of stars in main-sequence, it is difficult to accurately constrain the ages of these stars. Owing to the ambiguities inherent in estimating age, greater precision is required for these investigations. Unlike other stellar properties, such as $T_{eff}$, log$\:$g and [Fe/H], ages cannot be directly observed or measured. To estimate stellar ages, one uses an indirect model-dependent technique such as isochrone fitting \citep{2014EAS....65..225V}. Other approaches, such as gyrochronology and activity index, are also used in addition to isochrone fitting from stellar evolutionary models. Chemical analysis (also known as chemiochronology, \citep{2019A&A...624A..78D,2022AJ....164...60S}) and stellar kinematics \citep{2000AmJPh..68...95B,2021MNRAS.501.4917W} can be used to estimate the ages of an ensemble of stars but cannot be used for individual cases. Asteroseismology stands as the sole method capable of ascertaining the age of a star with an impressive level of precision, reaching uncertainties as low as $\sim$ 11$\%$ \citep{2019A&A...622A.130B}. However, it requires longer time-series data, which is only accessible for a limited sample of stars. Additionally, it only applies to stars hotter than about spectral type K, as cooler stars do not typically exhibit the oscillations required for estimating ages using asteroseismology \citep{2015MNRAS.452.2127S,2018haex.bookE.184C}. Each of the aforementioned models requires input parameters derived from various sources. Each input parameter is accompanied by its own uncertainty, which ultimately propagates into the age estimation.

\begin{table}
\caption{\label{t0} Sample distribution of stars hosting small and giant planets used in this paper.}
\centering
\begin{tabular}{lcc}
\hline\hline
Count &Stellar-hosts& Planets\\
%& (Gyrs) & (Gyrs) & \\

\hline
Total & 2336 & 3034 \\
Small ($M_{P}<0.3M_{J}$) & 1834 & 2509\\
Giant  ($0.3M_{J}\leq M_{P}\leq 5M_{J}$)& 502 & 526 \\
\hline
\end{tabular} \\
 Note -- The above values are listed after the sample curation as described in section~\ref{s2}. 

\end{table}

The ages of the individual planet-hosting stars are often determined using different models and input parameters. Due to the inherent uncertainties associated with each model and its input parameters, carrying out a meaningful statistical comparison becomes challenging. There have been limited homogeneous studies for the ages of the planet-hosting stars. Early studies such as \cite{2005A&A...443..609S} have estimated the stellar ages of the 49 planets hosting stars. Since the sample was very small, it was not possible to draw any robust conclusions about different populations of planets. Recent studies of 326 planets hosting stars by \cite{2015A&A...575A..18B} using PARSEC isochrones have found that $\sim$6$\%$ of stars have ages lower than 0.5~Gyr, while $\sim$7$\%$ of stars are older than 11~Gyr. Additionally, their results showed that the majority of their planet-hosting stars fall within the age range of 1.5 to 2 Gyr, indicating a prevalence of younger systems. Using astroseismology data for 33 Kepler exoplanet host stars, \cite{2015MNRAS.452.2127S} have found that the majority of these Kepler host stars are older than the Sun. Further study of 335 transiting planets hosting stars by \cite{2016A&A...585A...5B} has shown that the median age of the sample is $\sim$5~Gyr, which is similar to the solar age. These studies motivated us to look for possible correlations between stellar ages and planet mass. However, these studies are based on a limited sample of stars, and the ages derived in these papers have strong dependence  on the models and input parameters. As a result, the estimates become less reliable for studying any  statistical correlation between the stellar ages and the properties of their planetary companions.

\begin{figure}
\centering
\includegraphics[width=1\columnwidth]{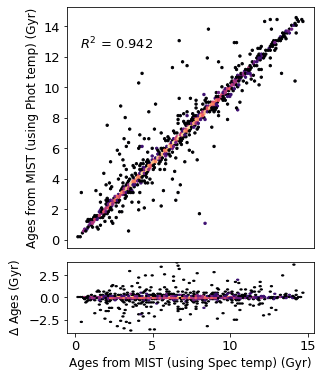}
\caption{Comparison of stellar ages computed from photometric $T_{\textrm{eff}}$ from Gaia DR3 and spectroscopic $T_{\textrm{eff}}$ from sweet-cat. The color coding represents the density of points.}
\label{f2}
\end{figure}

In order to understand how stellar ages are correlated with planetary properties, we need to analyse a large number of planet-hosting stars with minimal errors in their ages. Because of the difficulties in getting precise age estimates from isochrone models due to high uncertainties, we confirmed these estimates using established isochrone fitting models and parameters from photometric and spectroscopic data.
 Our findings indicate that, while slight variations in age estimates may occur based on the specific input parameters or models employed, the overall statistical trends for a large sample of planet hosting stars remain unchanged. In this study, our specific objective is to determine the ages of stars that host planets and examine the relationships between stellar ages and planetary properties, namely, the orbital period and mass of the exoplanets. We used combinations of input parameters from photometry and spectroscopy and three isochrone fitting models (MESA Isochrones \& Stellar Tracks (MIST)
 \citep{2016ApJS..222....8D,2016ApJ...823..102C}, PAdova and tRieste Stellar Evolutionary Code (PARSEC) \citep{2012MNRAS.427..127B} and q2-Yonsei-Yale (YY) \citep{2009gcgg.book...33H, Ramirez2014}) for the analysis. The paper is organized  as follows. In section~\ref{s2} we describe our sample of stars hosting planets. Section~\ref{s3} discusses the methodology of the age determination using isochrones. In section~\ref{s44} and section~\ref{s5} we discuss the trends for stellar ages as a function of planet mass and interpret the results in terms of the chemical evolution of our galaxy. Finally, we summarise and conclude our findings in section~\ref{s6}.

\begin{figure}
\centering
\includegraphics[width=1\columnwidth]{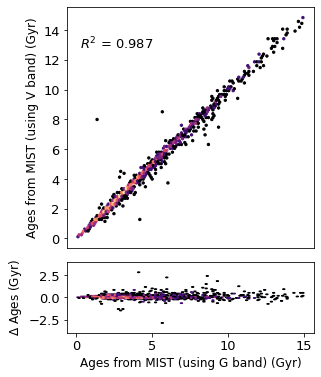}
\caption{Comparison of stellar ages from MIST model using  V-band (Johnson) and G-band (Gaia) mag. The color coding represents the density of stars.}
\label{f3}
\end{figure}

\begin{figure*}
\centering
\includegraphics[width=2.2\columnwidth]{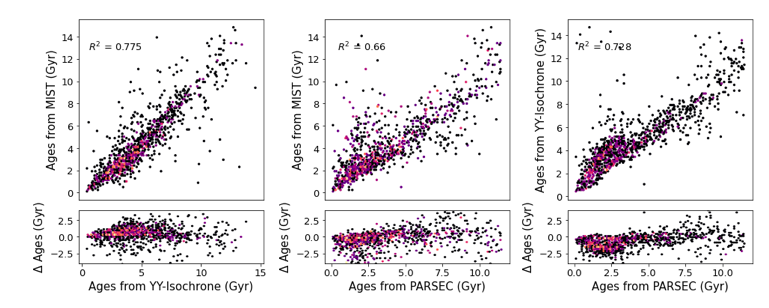}
\caption{Comparison of stellar ages from MIST, PARSEC and Yonsei-Yale isochrone fitting models. The color coding represents the density of stars.}
\label{f4}
\end{figure*}

\section{Sample selection}
\label{s2}
To study the dependence of stellar ages on planetary properties such as planet mass ($M_{P}$) and orbital period, we selected an initial sample of 3775 planet-hosting stars from the NASA exoplanet archive \citep{2013PASP..125..989A,https://doi.org/10.26133/nea12}. To calculate stellar ages using isochrone fitting techniques, we need accurate stellar parameters from spectroscopy (such as T$_{\textrm{eff}}$, log$\:$g and [Fe/H]), photometry (like G-band magnitude), and astrometry (like parallax, proper-motion etc). The  photometric and astrometric data was obtained from GAIA observations by cross-matched \footnote{Initially, we used a wider search radius, but in this instance, a 3" search radius was adequate to retrieve all planet-hosting stars. We also used other sources, such as the SIMBAD, to confirm that they are actual planet-hosting stars.} our sample with GAIA DR3 \citep{2022arXiv220800211G}, while for the spectroscopic data was obtained from the sweet-cat \citep{2017A&A...600A..69A,2018A&A...620A..58S,2021A&A...656A..53S} which is a catalogue of stellar parameters for SWP determined homogeneously from spectroscopy. We narrowed down our analysis to main sequence stars due to the complexities associated with accounting for evolutionary effects, such as photospheric mixing, which can introduce variations in photospheric metallicities \citep{2011JPhCS.328a2002B,2022AJ....164...60S}. Subsequently, we excluded super-Jupiters ($M_{P}\geq5M_{J}$) and multi-planetary systems hosting at least one small planet ($M_{P}<0.3M_{J}$) and a giant planet ($M_{P}\geq0.3M_{J}$), as including such multi-planetary systems would make it difficult to discern differences in stellar populations between small and giant planets as multi-planetary systems show properties similar to giant-planets and are statistically younger (see Appendix B of \cite{2022AJ....164...60S}). The distinction between small and giant planets is taken at the mass of Saturn, drawing on the mass-density rationale presented in \cite{2015ApJ...810L..25H}. This boundary is based on the observed shift in the slope within the mass-density relationship for exoplanets.Further, \cite{2017A&A...604A..83B} used Mass-Radius (M-R) relations and arrived at similar conclusions. They further suggested that the location of the breakpoint is linked to the onset of electron degeneracy in hydrogen, and therefore to the planetary bulk composition. For further analysis, we retained only those stars that have age uncertainties smaller than their main-sequence lifetime, as recommended by \cite{2004MNRAS.351..487P}. Additionally, we excluded lower main-sequence stars (T$_{\textrm{eff}}<$4400K) from our sample since the main-sequence lifetime for such stars can exceed the age of the universe, and therefore, the current stellar isochrone models cannot reliably estimate their age. Consequently, our final dataset comprised 2336 stars hosting 3034 planets (see table~\ref{t0}).

\section{Stellar age determination: Isochrones}
\label{s3}

To determine the ages from isochrone, one places the star on the Hertzsprung$-$Russell diagram (HRD) with $ T_{eff}$ on the x-axis and luminosity $L$ on the y-axis (Figure~\ref{f11}).
The $T_{eff}$ and  $L$ can be obtained by several techniques. In the case of $T_{eff}$, it can be determined both by spectroscopy and photometry (colour-index), while the $L$ is computed from the observed total flux, which is obtained from the photometric magnitude \citep{2017hsa9.conf..447R}, and distance from the parallax ($\pi$). These observables have their own intrinsic errors and systematics based on the techniques used to obtain them. Thus, the isochrone placement technique becomes challenging to determine the ages of a star with high accuracy. Further, for the lower main-sequence stars (T$_{eff}<$4400K), the isochrone ages are not reliable as the evolution timescales for these stars are $>$ age of the universe, and thus it becomes challenging to model the evolution of such systems.

Another factor that influences the determination of isochrone ages is the selection of models/grids. While different isochrone models share a common goal of estimating stellar ages, they diverge in their underlying assumptions. For instance, the equation of state (EOS) employed by the PARSEC models primarily relies on the Free EOS tool \footnote {https://freeeos.sourceforge.net/}. In contrast, MIST and YY isochrones predominantly utilize the OPAL (Opacity Project at Livermore) \citep{1996ApJ...464..943I} and the SCVH (Stewart, Colwell, Vasil, and Helfand) equation of state \citep{1995ApJS...99..713S}, respectively. Furthermore, variations in solar abundances among the different isochrone models contribute to discrepancies in their results. For example, YY-isochrones adopt solar abundances from \citep{1998SSRv...85..161G}, while MIST isochrones employ the values from \cite{2009ARA&A..47..481A}. Additionally, the choice of atmospheric models, such as ATLAS12, PHOENIX (BT-Settl), SYNTHE, MARCS, and others, in conjunction with the EOS, opacity values and solar abundances, further contributes to systematic differences in the estimation of stellar ages. These differences are critical factors that must be taken into account when determining isochrone ages. They highlight the complexities and uncertainties involved in age estimation and demonstrate the need for careful consideration and comparison of multiple isochrone models to mitigate potential biases.

We used isoclassify\footnote{https://github.com/danxhuber/isoclassify}\citep{2017ApJ...844..102H,2020AJ....159..280B,2023arXiv230111338B}, a robust tool designed to estimate stellar ages using MIST isochrone grids. For the estimation of stellar ages using PARSEC isochrones, we used PARAM-1.5 \footnote{http://stev.oapd.inaf.it/cgi-bin/param},  while for the Yonsei Yale (YY) isochrones, we used q2 tool \footnote{https://github.com/astroChasqui/q2}. To obtain the stellar ages, we used the observed stellar parameters such as effective temperature ($T_{\text{eff}}$), luminosity ($L$), and metallicity ([Fe/H]) as inputs into these codes, which then matches these observations with theoretical isochrones to estimate stellar ages. To ensure the robustness of our age determinations, we conducted a comparative analysis using the age estimates derived from each of the three isochrone sets. This approach allowed us to evaluate the consistency of age estimations across different stellar models and to identify any systematic discrepancies that may arise due to the underlying assumptions of each isochrone model.

Statistical inferences drawn for a sample of stars become inherently unreliable due to the dependence of individual star age determinations on both the input parameters and the models utilized. Therefore, we  estimate the ages of the stars using various models and different combinations of input parameters. Our objective is to assess whether consistent statistical conclusions could be drawn across different combinations of models and input parameters.  Here,  we vary certain input parameters (for instance, we use T$_{eff}$ form photometry and spectroscopy) while keeping the other parameters constant (for instance using G-band magnitude and MIST isochrone models) to demonstrate how the age varies from one case to another and how the overall stellar age for the population of exoplanets varies statistically.

\begin{figure*}
\centering
\includegraphics[width=2\columnwidth]{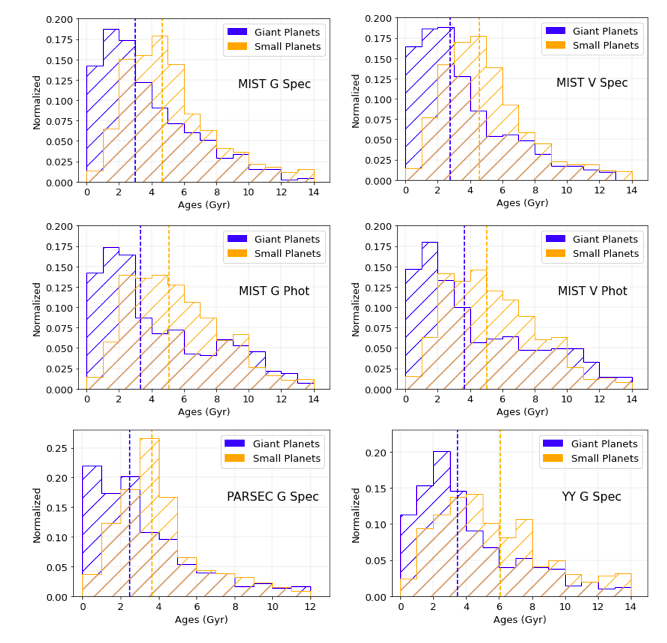}
\caption{Histogram for ages of the planets hosting stars for small and giant planets. The label is indicated in the following format: isochrone model-photometric band-temperature from spectroscopy or photometry. The dashed lines represent the median ages corresponding to their color labels in the histograms. }
\label{f5}
\end{figure*}
 
\begin{figure*}
\centering
\includegraphics[width=2\columnwidth]{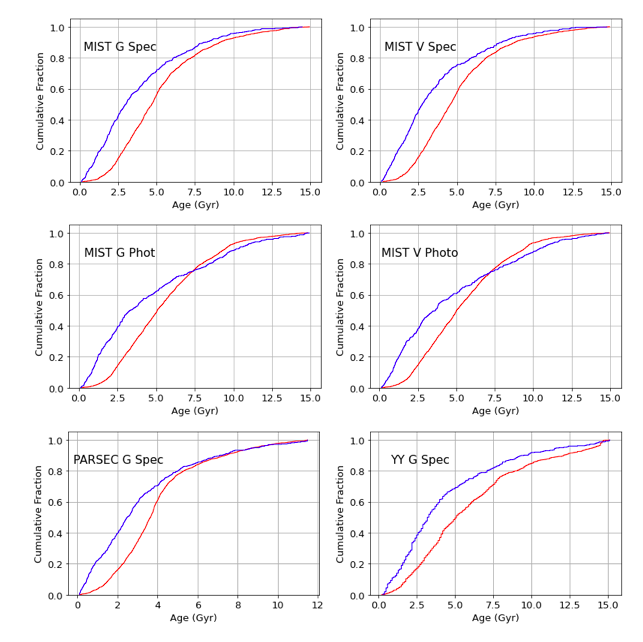}
\caption{Cumulative distribution for the stellar ages of small (red) and giant (blue) planet-hosting stars obtained using different isochrone models and input parameters.}
\label{f88}
\end{figure*}

\begin{figure*}
\centering
\includegraphics[width=2\columnwidth]{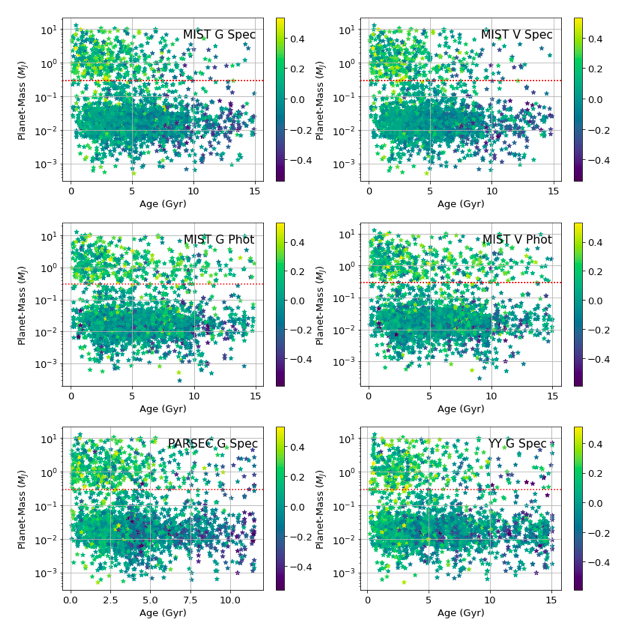}
\caption{Planet mass as a function of stellar age. The colorbar on the right represents the stellar metallicity values obtained from the literature. The dashed red line (M$_{P}$ = 0.3M$_{J}$) separates the small and giant planet-hosting stars. }
\label{f6}
\end{figure*}

\begin{table}
\caption{\label{t1}Comparison of stellar ages for the small and giant planet-hosting stars using different models and input parameters}
\centering
\begin{tabular}{lccc}
\hline\hline
&Giant-planet& Small-planet& p-value\\
& (Gyr) & (Gyr) & \\

\hline
MIST G Spec & 3.00$\pm$2.56 & 4.69$\pm$2.47 & $10^{-26}$\\
MIST G Phot & 3.34$\pm$3.12 & 5.08$\pm$2.95 & $10^{-16}$\\
MIST V Spec & 2.38$\pm$2.89 & 4.58$\pm$2.37 &  $10^{-16}$\\
MIST V Phot & 3.42$\pm$3.34 & 5.03$\pm$2.83 & $10^{-29}$\\
Yonsei Y G Spec & 3.45$\pm$3.15& 5.05$\pm$2.48  & $10^{-36}$\\
PARSEC G Spec & 2.63$\pm$2.21 &  3.78$\pm$2.01 &  $10^{-13}$\\
\hline
\end{tabular} \\
Note -- The errors quoted are the median absolute deviation (MAD) in the above distributions. Here the p-value obtained from the KS - test represents the probability that the two samples belong to the same distribution.

\end{table}

\subsection{Choice of stellar temperatures}
The temperature of a star is mostly obtained by spectroscopy or photometric measurements. Both of these techniques have certain assumptions while obtaining the estimate of the $T_{eff}$ and this leads to a systematic difference in the estimation of ages \citep{1979RA......9..519W}. To verify how the input ages affect the stellar age estimates from isochrone fitting we took the $T_{eff}$ from two sources: sweet-cat \citep{2021A&A...656A..53S} (spectroscopic) and GAIA DR3 \citep{2022arXiv220800211G} (photometric). Figure~\ref{f2} shows the spread in age using photometric and spectroscopic temperatures using Yonsei Yale isochrone models \citep{2009gcgg.book...33H} and GAIA parallaxes \citep{2021A&A...649A...4L}. We find that $\sim$85$\%$ of the stars have an age difference $<$ 0.5 Gyrs. Similarly, when employing MIST and PARSEC isochrone models, we observe that 72$\%$ and 75$\%$ of the stars, respectively show an age difference of less than 0.5 Gyrs, indicating that the scatter is small with no significant systematic in the age estimates from spectroscopic and photometric temperatures.

\subsection{Choice of photometric band magnitude}
The luminosity estimate in our model relies on the total flux ($f$), which is determined from the band magnitude provided. Given this, we explored whether the selection of photometric band magnitudes has any noteworthy impact on the process of estimating stellar ages. We decided to use the Johnsons V-band and Gaia G-band magnitudes to estimate the ages of the exoplanet host star. Stars for which V-band magnitude was not available, we used the relationship as described in GAIA archive \footnote{https://www.cosmos.esa.int/web/gaia-users/archive/gdr3-documentation} and used G, G$_{bp}$ and G$_{rp}$  magnitudes to obtain the V-band magnitude. We also tested the above relation for the stars whose both G-band and V-band were available and we found the V-band obtained from the empirical relation matches with the observed ones. Figure~\ref{f3} shows the correlation between the ages determined from V- and G-band magnitudes using spectroscopic temperatures and MIST. We find that the ages obtained from G-band and V-band are strongly correlated and do not show any significant systematic differences. We also performed this analysis with several other combinations of input parameters (for instance, using photometric and spectroscopic $T_{eff}$ ) and did not find any considerable dispersion in any case.

\subsection{Choice of models}
Most of the stellar ages obtained using isochrone use a standard stellar evolutionary isochrone fitting model to estimate the age of stars. However, the choice of the model plays an equally important role just as input parameters \citep{2019A&A...624A..78D}. Figure~\ref{f4} shows the scatter plot in the stellar ages obtained using various isochrone fitting models using spectroscopic temperatures and G magnitude. We find that the ages estimated using MIST and YY are well in agreement, with a moderate spread but no significant systematic differences. For the stellar ages obtained using PARSEC models, we find a large scatter for stars with ages $>$6~Gyr, when comparing with the ages obtained using MIST and YY. We also find a notable systematic difference in the stellar ages obtained from PARSEC models when comparing the ages with MIST and YY.

\begin{table*}
\caption{\label{t2}Distribution of Jupiter hosting stars in terms of their stellar ages and orbital period.}
\centering
\begin{tabular}{lccccccc}
\hline\hline
& Young hot Jupiters & Old hot Jupiters & Young cool Jupiters & Old Cool Jupiters & Hot Jupiters & Cool Jupiters & p-value\\
& (Count) & (Count)& (Count) & (Count) &(Gyr) & (Gyr) \\

\hline
MIST & 207 & 60 & 115 & 46 &2.43 &3.50 & $10^{-3}$\\
%PARSEC & 3.85 & 2.67 & 2.58 x $10^{-11}$\\
%\textbf{MIST} V Spec & 3.66 & 2.51 & 3.11 x $10^{-15}$\\
%\textbf{MIST} V Phot & 3.80 & 2.70 & 9.91 x $10^{-12}$\\
Yonsei Y & 191 & 76 & 97 & 67 & 3.08 & 4.35 & $10^{-5}$\\
PARSEC  & 289 & 48 & 132 & 56 & 2.15 & 3.38 &  $10^{-9}$\\
\hline
\end{tabular} \\
The standard error of the mean is typically 0.01 Gyr for all the cases. The p-value is obtained in the same way as in table~\ref{t1}.

\end{table*} 
% \subsection{Choice of parallax}
% To estimate how the choice of parallax affects the stellar age estimates, we decided to compare the ages determined from Hipparcos parallaxes \citep{1997A&A...323L..49P} and those obtained from Gaia DR3 parallaxes \citep{2022arXiv220800211G}. We find that the errors in parallax measurements are better by a factor of two for the Gaia values. For a test case, we used the PARSEC isochrone models \citep{2012MNRAS.427..127B} with spectroscopic temperatures to estimate the stellar ages. Although the individual age measurements do not vary significantly, the error estimates are well constrained for the Gaia parallax values. Since, Gaia DR3 parallaxes are the most accurate so far, we used the Hipparcos parallax as a simple case study on how the parallax measurement plays a role in estimating the stellar ages. We didn't use hipparcos parallaxes for any further statistical analysis in this paper.

\section{Results}
\label{s44}
\subsection{Ages of the planet-hosting stars}
We compute the ages of the planet-hosting stars from isochrone fitting methods using different models and input parameters. Figure~\ref{f5} shows the age histograms of stars hosting small planets and giant planets. Note that,  even though the distribution of stellar ages computed using different models and input parameters has noticeable differences, the relative offset (median age difference) between the ages of small and giant planet-hosting stars, is always positive. In other words, the median ages of the stars hosting small planets are higher compared to stars hosting giant planets in all cases (see Table~\ref{t2}).  We also find that the median age of stars obtained from PARSEC models is slightly lower when compared with the MIST and YY isochrone models, and this is due to the systematics of the stellar ages as obtained in Section~\ref{s3}.

Further, the cumulative age fraction shown in figure~\ref{f88} implies that stars hosting small and giant planets belong to different populations. We also performed the Kolmogorov Smirnov (K-S) test on the sample and found that the sample of stars hosting small and giant planets fall into distinct age groups (table~\ref{t1}). This result clearly suggests that small planets are common around both young and older stars, whereas giant planets are more prevalent around younger stars.

It is important to note that due to high stellar activity and RV jitter\footnote{RV jitter is the intrinsic noise in radial velocity measurements of a star, caused by factors such as stellar activity, granulation, oscillations, and instrumental limitations. It poses challenges in detecting planets, raising the detection threshold, and introducing false positives/negatives.}, the detection of small and low-mass planets around young stars is far more challenging than the detection of giant planets. Therefore, in the current exoplanet census, there is a strong possibility that some small planets might not have been detected around younger stars. However, the lack of giant planets around older stars is not likely caused by any detection or selection bias, as detecting giant planets is relatively easier than detecting small planets, regardless of the method used or the age of the star. 

\subsection{Planet mass as a function of stellar age}
Figure~\ref{f6} shows the planet-mass distribution as a function of stellar age with $\sim$ 70\%-85\% of stars in the sample having an age below 7~Gyr. We also note that the population of giant planets began to rise about 4-5~Gyr ago, indicating that the formation epoch of giant planets is relatively recent compared to the population of small planets, with the formation onset occurring as early as 7-8 Gyr ago. Moreover, the colorbar in Figure~\ref{f6}, which signifies stellar metallicity (taken from Sweet-Cat \cite{2021A&A...656A..53S}), reveals a metallicity gradient. This suggests that young stars that host planets are statistically richer in metals. This is also in line with the previous studies that have shown giant planet-hosting stars are metal-rich \citep{fis05,sou19,nar18,2022AJ....164...60S,2023AJ....166...91S}, which supports the core-accretion model for planet formation.

\begin{figure}[t]
\centering
\includegraphics[width=1\columnwidth]{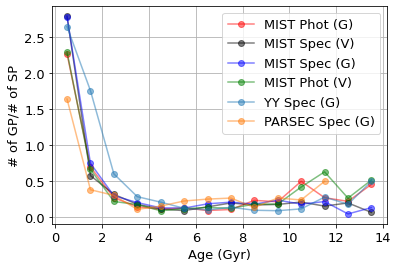}
\caption{Fraction of stars hosting giant planets to small planets as a function of stellar age using different models and input parameters. The data is binned at an interval of 1~Gyr.}
\label{f7}
\end{figure}
\begin{figure*}[t]
\centering
\includegraphics[width=2.2\columnwidth]{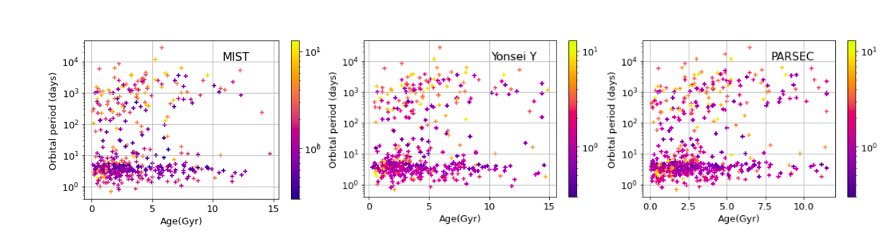}
\caption{Distribution of giant planets as a function of stellar age. The colour bars represent planet mass and labels in each plot window represent the isochrone models used. For all three cases, spectroscopic $T_{eff}$ and G magnitude were used as input parameters.}
\label{f8}
\end{figure*}

\subsection{Planet fraction vs stellar age}
In order to understand the progression of planet formation in the galaxy, we calculated the ratio of stars hosting giant planets to the stars hosting small planets as a function of the stellar age. In all cases that we analyzed in this paper, the ratio falls as a function of stellar age, as shown in figure~\ref{f7}. However, It is possible that the ratio is higher for younger stars ($<$5~Gyrs) due to selection biases and sensitivity limitations of current detection techniques.  However, for stars, $>$5~Gyrs, the finding is certainly not due to any selection or observation biases. This is because regardless of stellar type, the giant planets are easier to detect with any technique compared to the small planets. This finding possibly suggests that it is more likely that young stars have a higher ratio of giant planets to small planets when compared to older stars. A small scatter at the age bin 10-12~Gyr in figure~\ref{f7} is due to the small number of stars in that bin.

\subsection{Hot Jupiters are younger}
For a subsample of the giant planet population, we also looked at the distribution of their orbital period as a function of the ages of their parent stars obtained from different models using spectroscopic $T_{eff}$ and G-band magnitude. A closer look at figure~\ref{f8} shows that the majority of these stars are young ($\leq$ 5 Gyrs) and they are hosting hot-Jupiters with orbital periods of fewer than 10 days. This is evident from the clustering of points around the lower-left sides of each plot in figure~\ref{f8}.  For comparison, the median age of hot-Jupiters (orbital period $\leq$ 10 days) and warm/cool Jupiters (orbital period $>$~10~days) is listed in table ~\ref{t2}. On average, the hot Jupiter hosting stars are $\sim$ 1.2~Gyr younger than the stars harboring warm/cool Jupiters.

Specifically, we note that $\sim$~70\% of giant planets have age 5~Gyr or less, indicating that overall the giant planetary systems are younger. Although the high number of young hot Jupiter (lower-left region of each plot in figure~\ref{f8}) can be possibly due to detection bias as hot Jupiters are easier to detect \citep{2016MNRAS.463.1323K}, we find that the number of young warm/cool Jupiters  (top-left region of each plot in figure~\ref{f8}) is notably higher than old hot-Jupiters (bottom-right region of each plot in figure~\ref{f8}), which is unlikely due to any observational bias. In fact, it again points to the scenario of the late onset of giant planet formation in the galaxy.

\subsection{Isochrone vs. Asteroseismology and Chemical Clock Ages
}
\begin{figure}[t]
\centering
\includegraphics[width=1\columnwidth]{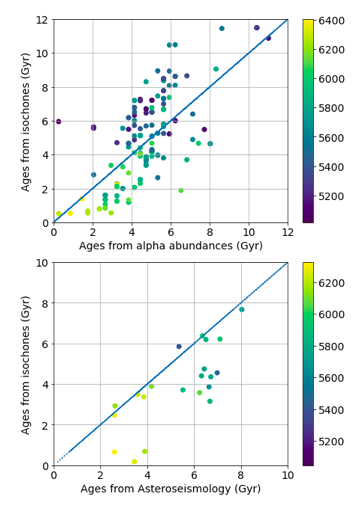}
\caption{\textbf{Top: }Comparison of ages obtained from stellar isochrones and from $[\alpha$/Fe] abundances. \textbf{Bottom:} Comparison of ages obtained from stellar isochrones and from Asteroseismology. The colour represents the effective temperature of the star. }
\label{f99}
\end{figure}

We compared the stellar ages derived from MIST isochrone models with those obtained using Asteroseismology and from $\alpha$ abundances (also known as chemical clocks). For Asteroseismology we used the ages from \cite{2015MNRAS.452.2127S} and our analysis reveals that isochrone-based ages are slightly overestimated when compared to asteroseismology-derived ages as shown in Figure~\ref{f99}. Conversely, when comparing the ages obtained using the $\alpha$ abundances using the relationship in \cite{2019A&A...624A..78D}, we find that the ages are over-estimated for the younger stars ($\leq$ 4 Gyrs) while they slightly underestimated for the older stars ($>$ 4 Gyrs) as shown in Figure~\ref{f99}.

The discrepancies observed between the ages derived from MIST isochrone models, asteroseismology, and $\alpha$ abundances underscore the need for methodological refinement across age-determination techniques. Specifically, the tendency of asteroseismology to underestimate ages, compared to isochrone models, and the accuracy of chemical clocks for stars of different ages highlights the importance of cross-validating stellar ages to identify and correct systematic biases.

\section{Discussion}
\label{s5}
The results from section 4 indicate that the ages for the majority of stars hosting planets are around $~$6 to 7 Gyrs depending on the model. Further, within the planet's population, the giant planet-hosting stars are younger when compared to the stars hosting small planets. The reason why younger stars host more giant planets compared to older stars could be understood from the knowledge of the chemical composition and dust-to-gas ratio of their circumstellar environment. A higher dust-to-gas ratio favors the formation of giant planets \citep{2015A&A...582L..10K} as:  a) the solid dust grains act as the building blocks for planetesimals and planetary cores. When the dust-to-gas ratio is higher, it means there is a larger amount of solid material available compared to the surrounding gas. This increased availability of material provides a larger reservoir for the growth and accumulation of solid cores. b)  In a higher dust-to-gas ratio environment, collisions between dust grains become more frequent. These collisions can lead to sticking and aggregation, allowing the particles to grow in size. With a greater number of collisions occurring, the growth process can proceed more rapidly, enabling the formation of larger planetesimals and planetary cores over shorter timescales ($\sim$ 5-10 Myrs). c) Larger solid cores have stronger gravitational forces, enabling them to attract and capture more surrounding gas. When the dust-to-gas ratio is higher, there is a denser population of dust grains that can coalesce into larger planetesimals and cores. These more massive cores can then more effectively accrete gas from the protoplanetary disk, rapidly increasing their size and leading to the formation of gas giant planets \citep{2021A&A...656A..70E,2022arXiv220309759D}.

Together with the dust-to-gas ratio, the grain composition also plays a key role in the formation of the planetary core \citep{1995A&A...300..503D,2001A&A...378..228F,2003ApJ...598.1017D}. During the early stages of galaxy evolution, the circumstellar disk is devoid of sufficient grains as the ISM is mostly enriched with Type II supernovae. The dust grains consisted of mostly silicates, magnesium, and other $\alpha$ but lacked iron or other heavier elements, thus resulting in a lower dust-to-gas ratio ($<<$0.01). As the galaxy evolved, the ISM was enriched with Fe-peak elements from Type Ia supernovae, which in turn increased the dust-to-gas ratio and thus favouring the formation of both small and giant planets \citep{2015A&A...579A..52N,2018ApJ...865...68B,2018A&A...619A.125A,2018MNRAS.477.2326F,2019A&A...624A..19B,2019A&A...624A..78D}. The presence of giant planets around metal-rich and young stars also correlates with higher dust-to-gas ratio ($\sim$ 0.01) in young and metal-rich protoplanetary disks which makes them conducive to the formation of giant planets. This is also consistent with the core-accretion process \citep{1996Icar..124...62P,2007ApJ...662.1282M} leading to the formation of giant planet, where a solid core of $\sim$ 10-15 M$_{\earth}$ needs to form quickly ($\sim$ 10 Myr) before the gas in the disk dissipates \citep{2001ApJ...553L.153H,2012ApJ...745...19K}, otherwise the resulting planet would end up rocky in nature. The presence of a higher dust-to-gas ratio, thus promotes faster core formation, thereby facilitating the formation of giant planets.

\section{Summary and conclusions} 
\label{s6}
The characteristics of exoplanets and their host stars exhibit a close interdependence. In the present study, we focused on estimating the stellar ages of planets orbiting main-sequence stars. To accomplish this, we used the isochrone fitting technique to estimate the stellar ages of the planet-hosting stars. Furthermore, we conducted an extensive analysis, exploring possible correlations between stellar and planetary properties to gain insights into their formation mechanisms. Our sample consisted of 2336 stars known to host planets, detected through both transit and radial velocity (RV) methods. In conclusions:

\begin{enumerate}

\item We computed the stellar ages for the main sequence planets hosting stars using the isochrone fitting technique. Since isochrone ages are highly model- and input-parameter-dependent, we used several models and input parameters in order to understand the systematic age differences obtained from different isochrone models.
\item We find that, even though individual age estimates and their distributions vary depending on the choice of model or input parameters, the underlying statistical trends remain unaffected.

\item Our findings suggest that ~70\% to 85\% of planets have stellar ages $<$ 7Gyrs and most of the planets started forming after the ISM was enriched sufficiently to form the cores of the planets.
\item Our analysis reveals a distinct divergence in the ages of stars hosting small planets compared to those hosting giant planets. Specifically, we observe a statistically significant age difference, with stars hosting giant planets being notably younger than those hosting small planets. This disparity suggests that the formation of Jupiter-sized planets occurred at a later stage in the galaxy's evolution, specifically when the necessary dust-to-gas ratio had reached a threshold, enabling the formation of a significant number of giant planets. These findings corroborate the core-accretion theory of planet formation.
% \item The stars hosting small and giant planets belong to different classes of population in terms of their ages. We also performed a K-S test and found that stars hosting small and giant planets are significantly different.
% \item We find that the fraction of stars hosting small and giant planets falls rapidly as a function of stellar age, with the ratio saturating after 4 Gyr. 
\item Among the giant-planet population, we find that the hot Jupiters are the youngest, and they are the most recently formed systems in the context of planet formation.
\end{enumerate}

In conclusion, we have analyzed the stellar ages for a large number of exoplanet-hosting stars, connecting the planet formation process to the ages of their hosts. The fact that stars hosting giant planets are younger is largely consistent with the chemical evolution of the galaxy. From the observed trends between stellar ages and planet masses, we conclude that the small planet formation started to rise after the ISM was sufficiently enriched ($\sim$ 6-7 Gyr), while the giant planet formation is much younger and has started to form in large numbers only in the past $\sim$ 4-5 Gyr. 

%The major challenge in estimating the ages of the stars using the isochrone method is the large uncertainties in its measurements. To strengthen the findings that are presented in this paper, future astroseismology observations on a larger sample of exoplanet-hosting stars might shed some light, though astroseismology has its own limitations and is only applicable to specific groups of stars. Future stellar age estimates from various techniques such as stellar kinematics and chemical cartography might also shed some light on the correlations between stellar ages and planet masses.\\

\section*{ACKNOWLEDGMENTS}
This work has made use of (a) the NASA Exoplanet Database, which is run by the California Institute of Technology under an Exoplanet Exploration Program contract with the National Aeronautics and Space Administration and (b) the European Space Agency (ESA) space mission Gaia, the data from which were processed by the Gaia Data Processing and Analysis Consortium (DPAC) (c) the exoplanet.eu database maintained developed and maintained by the exoplanet TEAM. C Swastik would also like to thank Athul Ratnakar for the insightful discussion.

\software{Numpy \citep{2020Natur.585..357H}, Topcat \citep{2005ASPC..347...29T}, Astropy \citep{astropy:2013}, Scikit-learn \citep{scikit-learn}, Matplotlib \citep{4160265}, Scipy \citep{2020NatMe..17..261V} }

\bibliography{biblio}{}
\bibliographystyle{aasjournal}

% \newpage
% \appendix
% \section{NEW}
% \begin{figure*}[t]
% \centering
% \includegraphics[width=1\columnwidth]{COMP.png}
% \caption{Cumulative distribution for the stellar ages of small and giant planet hosting stars.}
% \label{f8}
% \end{figure*}
\end{document}